\documentclass[10pt,journal,compsoc]{IEEEtran}
\usepackage{times}  
\usepackage{helvet}  
\usepackage{courier}  
\usepackage{url}  
\usepackage{graphicx}  
\usepackage{multirow}
\usepackage{comment}
\usepackage{adjustbox}
\usepackage{color,soul}
\usepackage{authblk}

\title{Embedding-based Qualitative Analysis of Polarization in Turkey}

\begin{document}

\author[1] {Mucahid Kutlu}
\author[2] {Kareem	Darwish}
\author[3] {Cansin	Bayrak}
\author[4] {Ammar	Rashed}
\author[5] {Tamer Elsayed}
\affil[1,3]{TOBB University of Economics and Technology, Ankara, Turkey}
\affil[2]{Qatar Computing Research Institute,  Doha, Qatar}
\affil[4]{Ozyegin University, Istanbul, Turkey}
\affil[5]{Qatar University, Doha, Qatar}
{
    \makeatletter
    \renewcommand\AB@affilsepx{: \protect\Affilfont}
    \makeatother

    \affil[ ]{Email}

    \makeatletter
    \renewcommand\AB@affilsepx{, \protect\Affilfont}
    \makeatother
    \affil[1]{m.kutlu@etu.edu.tr}
    \affil[2]{kdarwish@qf.org.qa}
    \affil[3]{c.bayrak@etu.edu.tr}
    \affil[4]{ammar.rasid@ozu.edu.tr}
    \affil[5]{telsayed@qu.edu.qa}
}

%



\maketitle              

\begin{abstract}
On June 24, 2018, Turkey conducted a highly-consequential election in which the Turkish people elected their president and parliament in the first election under a new presidential system. During the election period, the Turkish people extensively shared their political opinions on Twitter. One access of polarization among the electorate was support for or opposition to the reelection of Recep Tayyip Erdo\u{g}an. 
In this paper, we explore the polarization between the two groups on their political opinions and lifestyle, and examine whether polarization had increased in the lead up to the election. We conduct our analysis on two collected datasets covering the time periods before and during the election period that we split into pro- and anti-Erdo\u{g}an groups. For the pro and anti splits of both datasets, we generate separate word embedding models, and then use the four generated models to contrast the neighborhood (in the embedding space) of the political leaders, political issues, and lifestyle choices (e.g., beverages, food, and vacation). Our analysis shows that the two groups agree on some topics, such as terrorism and organizations threatening the country, but disagree on others, such as refugees and lifestyle choices. Polarization towards party leaders is more pronounced, and polarization further increased during the election time.
\end{abstract}


%
%
%
\section{Introduction}
On June 24, 2018, Turkey conducted early elections for the presidency and the parliament that would bring into force the constitutional changes that were approved by referendum on April 16, 2017. The constitutional changes would transform Turkey from a parliamentary system to a presidential system. With the office of the president enjoying significantly-increased power, these elections were considered highly consequential for Turkey. Nascent coalitions were formed in the lead up to the election with several presidential candidates representing different Turkish political blocks such as conservatives, secularists, nationalists, and Kurds. Given the front runner status of the incumbent candidate, Recep Tayyip Erdo\u{g}an, and the newly-formed alliance between his party (AKParti) and the nationalist party (MHP), we explore  the polarization between those who supported Erdo\u{g}an and those who favored the other candidates in this work.

For our analysis, we collected 108M election-related tweets between April 29 and June 23, 2018. Then, using a semi-automatic labeling (based on self-declarations in users' profiles followed by a label propagation method), we labeled about 652.7K Twitter users, of which 279.2K are pro-Erdo\u{g}an and 373.5K are anti-Erdo\u{g}an. We estimate that tagging accuracy is above 95\%. Of those users, we crawled the timelines of 82K and 86K random users from pro- and anti-Erdo\u{g}an groups respectively to obtain tweets that were posted before and after the election. Timeline crawling yielded 213M tweets.

Previous work has suggested that word embeddings can capture human biases \cite{caliskan2017semantics} and hence determine shifting and/or divergent attitudes \cite{garg2018word,giatsoglou2017sentiment}.  We built upon this to conduct our analysis of polarization based on stem-based embeddings on the election-related and timeline tweets of pro- and anti-Erdo\u{g}an users.
Looking at both types of collections allowed us to observe polarization in general and during the election in specific, and to determine if polarization is transient or more systemic and whether it is pragmatic or ideological. We queried the trained embeddings to identify the nearest words to entities of interest in the embeddings space. We contrasted how different groups regarded prominent politicians, what positions they take on popular political issues, and which lifestyle choices, such as food and leisure, they make. Given the nearest words to entities of interest, we examined the relative rankings: of positive and negative sentiment words; words indicative of stance; and words highlighting lifestyle choices. Since we were training and querying embeddings on Turkish tweets, we applied Turkish and tweet-specific pre-processing such as normalization, stemming, and partial word matching to overcome ubiquitous spelling mistakes. We also trained embeddings that are based on sub-word character n-grams to overcome the morphological complexity of Turkish and users' creative spellings in tweets.  

Our contributions in this work are three-fold:
\begin{itemize}
    \item We collected a large collection of tweets related to the Turkish election containing more than 108M tweets, and used semi-supervised methods to \emph{accurately} tag more than 652k users given limited manual-tagging. We then collected more than 213M tweets from timelines of users from pro- and anti-Erdo\u{g}an groups.
    \item We used word embeddings to \emph{qualitatively} study polarization in the context of Turkish politics. Specifically, we employed stem-based embeddings based on the tweets of users with opposing views to ascertain users attitudes towards politicians and political issues as well as their lifestyle choices.
    \item We applied pre-processing techniques that are suitable for handling the morphological and orthographic specificities of Turkish and the ubiquitous misspelling in Tweets.
\end{itemize}

\section{Related Work}
\paragraph{Stance Detection} 
Stance detection can be performed 
using supervised classification and using a variety of features such as text-level features (e.g., words or hashtags), user-interaction features (e.g., user mentions and retweets), and profile-level features (e.g., name and location) \cite{borge2015content,magdy2016isisisnotislam,magdy2016failedrevolutions}. 
The use of retweets seems to yield competitive results \cite{magdy2016isisisnotislam,wong2013quantifying,wong2016quantifying}. 
Label propagation is also an effective semi-supervised method that propagates labels in a network based on follow or retweet relationships \cite{borge2015content,weber2013secular} or the sharing of identical tweets \cite{darwish2018scotus,kutlu2018devam,magdy2016isisisnotislam}. In this paper, we use an iterative label propagation method based on retweeted tweets. Other methods for user stance detection include: collective classification \cite{duan2012graph}, where users in a network are jointly labeled, and projecting users into a lower dimensional user space prior to classification \cite{darwish2017improved}.  More recent work projects user onto a two dimensional space then uses clustering to perform unsupervised stance detection \cite{darwish2019unsupervisedStance}.  
\paragraph{Polarization on Twitter} 
\label{sec:quantifyingPolarization}
Social media is a fertile ground for polarization, due to two social phenomena, namely: homophily, which is the tendency of similar users to congregate together, and biased-assimilation, where individuals readily accept evidence confirming their group's view, but are rather critical when provided with disconfirming evidence. Both phenomena are amplified on social network platforms \cite{bias_assim,homophily,homophily_twitter}.
Online social networks facilitate discovery and communication between like-minded users and hence the creation of large homophilous communities \cite{homophily_twitter}. Biased-assimilation has been shown to play a crucial role in the dynamics of polarization, as it makes community members more entrenched in their views, particularly for controversial topics \cite{bias_assim}. 
The dynamics of \textit{intra-community} and \textit{inter-community} interactions provide predictive information about potential conflicts \cite{weber2013secular}. Kumar et al. \cite{kumar2018community} introduced a method that employs graph embeddings, where the graph captures user interactions on Reddit\footnote{https://www.reddit.com/}, in conjunction with user, community, and text features to predict potential conflict and subsequent community mobilization. Unlike Twitter, the pseudo-anonymous nature of Reddit users may affect the types of interactions between communities. 
Other work has focused on quantifying polarization \cite{garimella2018quantifying,guerra2013measure,morales2015measuring}.  Several methods were used including random graph walks, network betweenness,   distances in embedding spaces for different groups \cite{garimella2018quantifying}, inter-group and intra-group distances \cite{morales2015measuring}, and popularity of boundary nodes between communities \cite{guerra2013measure}. Given polarized communities, several studies looked at identifying distinguishing features, such as hashtags, between such communities. One method uses the so-called valence score that measures the relative probability of a feature appearing in one community compared to another \cite{conover2011political,darwish2018scotus,weber2013secular}. Others have trained word embeddings, based on the content generated by different communities, to contrast them through the usage of similar words and words associated with different concepts \cite{garg2018word,giatsoglou2017sentiment}. Word embeddings were shown to capture implicit human biases \cite{caliskan2017semantics}. Garg et al. \cite{garg2018word} trained temporal word embeddings that span 100 years to measure shifts in racial and gender attitudes. They also correlated key concepts with positive and negative adjectives over time.  We extend their work in several directions, namely: we apply their method to polarized communities; we perform tweet-specific and Turkish-specific adaptations, such as normalization, stemming, approximate word-matching, and training of sub-word character n-gram embeddings; and we employ several variations of the technique to identify other notable aspects such as lifestyle preferences. 
Giatsoglou et. al. \cite{giatsoglou2017sentiment} trained a polarity classification model using word embeddings with a seed lexicon of polarity-labeled words. We utilize user labels to avoid such manual lexicon construction. We show that capturing contrasting polarities of words in different contexts reflects the general stance of the authors toward a topic.



\section{Background: Turkish Elections}

Through a referendum on April 16, 2017, Turkey made significant changes in its constitution, effectively changing the government from a parliamentary system to a presidential system, giving more power to the president. 
the Turkish president Erdo\u{g}an announced the first election under the new constitution to be held on June 24, 2018 to elect both the president and parliament simultaneously. 
Voter participation was 86.24\% with eight political parties participated in the parliamentary elections.  Table~\ref{tab:politicalParties} lists the main political parties 
and their election results (ignoring the minor political parties)\footnote{\url{http://ysk.gov.tr/doc/dosyalar/docs/24Haziran2018/KesinSecimSonuclari/2018MV-96C.pdf}}. 
For the first time in Turkish elections history, parties were also allowed to make alignments for parliamentary elections, such as the ``Public Alignment'', which included AKParti and MHP, and the ``Nation Alignment'', which included CHP, IYI and SP. Such brought parties with different ideological background together. In the presidential elections, there were 5 candidates from these major parties, namely Recep Tayyip Erdo\u{g}an (AK Parti), Muharrem Ince (CHP), Selahattin Demirta\c{s} (HDP), Meral Ak\c{s}ener (IYI), and Temel Karamollao\u{g}lu (SP). 
MHP and Huda-Par (a minor Kurdish Islamist Party) announced their support for Erdo\u{g}an in the presidential election. 

The incumbent and front runner status of Erdo\u{g}an caused voters to cast the elections as referendum to allow him to continue his presidency or not. Hence the hashtag  \#devam (meaning ``continue'') became popular among his supporters, while his opponents (regardless of their political affiliation) used the hashtag \#tamam (``enough''). In our data also, we have seen that many users used these hashtags, not just in their tweets, but also in their screen names. Therefore, Turkish voters can be roughly divided into two groups: pro and anti-Erdo\u{g}an voters.

We have also observed this political binarization in Turkish politics in the last elections held on 31 March, 2019. 
CHP and IYI party again formed Nation Alignment while MHP and AK Parti formed Public Alignment.  Even though HDP was not a part of the Nation Alignment and did not field candidates for many cities, HDP announced support for the most favorable candidate running against the Public Alignment's candidate in such cities\footnote{https://www.toplumsal.com.tr/hdp-neden-3-buyuksehirde-aday-gostermedi-pervin-buldan-acikladi/}.    


\begin{table*}[ht]
\caption{Main political parties that participated in the parliamentary elections on June 24, 2018 along with their election results. Abbreviations are used based on their official Turkish name.}
\centering
\begin{tabular}{ l| l |c |l } 
\textbf{Party Name} &  \textbf{Description} & \textbf{Alliance} & \textbf{Election Results} \\  \hline
Justice \& Development (AKParti) & Erdo\u{g}an's party & Public & 42.56\% \\ \hline 
Nationalist Movement (MHP)& Turkish nationalist party  & Public & 11.1\% \\ \hline \hline
Republican People's (CHP) & Secular party founded by Ataturk & Nation & 22.64\% \\ \hline
Good (IYI) & Turkish nationalist party founded by mostly ex-members of MHP & Nation & 9.96\% \\ \hline
Saadet (felicity) (SP) & Islamist Party & Nation & 1.34\% \\ \hline
People's Democratic (HDP) & Secular Left-wing Kurdish party & None & 11.7\% \\ \hline
		\end{tabular}
		\label{tab:politicalParties}
\end{table*}

\section{Dataset}
We constructed two different datasets in our study. First, we collected election-related tweets. Next, we labeled the users as pro-Erdo\u{g}an (Pro) or anti-Erdo\u{g}an (Anti) using a two-step label-propagation approach. The tweets of labeled users constitute our  Election Dataset (ED). Subsequently, we crawled all tweets of randomly selected labeled users to construct our Timeline Dataset (TD).
\paragraph{\textbf{Election Data Crawling:}}\label{section_election_time_crawling}
We collected tweets related to Turkey and the election starting on April 29, 2018 until June 23, 2018, which is the day before the election. We tracked keywords related to the election including political party names, candidate names, popular hashtags during this process (e.g., \#tamam and \#devam), famous political figures (e.g., Abdullah G\"{u}l, the former president of Turkey), and terms that may impact people’s vote (e.g., economy, terrorism and others). We wrote keywords in Turkish with Turkish alphabet, which contains some additional letters that do not exist in the English alphabet (e.g., \c{c}, \u{g}, \c{s}). Next, we added versions of these keywords written strictly with English letters (e.g., ``Erdo\u{g}an'' instead of ``Erdo\u{g}an'') allowing us to catch non-Turkish spellings. Overall, we collected 108M tweets. 
\paragraph{\textbf{Labeling:}}\label{sec_labeling} 
The labeling process was done in two steps:

\textbf{(1) Manual labeling:} First, we assigned labels to users who explicitly specify their party affiliation in their Twitter handle or screen name. We made one simplifying assumption, namely that supporters of a particular party would be supporting the candidate supported by their party. We extracted a list of users who use ``AKParti", ``CHP", ``HDP", or ``IYI" in their Twitter user or screen name. We labeled the users who used ``AKParti” as ``pro-Erdo\u{g}an”, while the rest as ``anti-Erdo\u{g}an”. Though ``MHP” officially supported Erdo\u{g}an in the election, we feared that the MHP supporters might not be universally supporting Erdo\u{g}an.
Further, 
we labeled users who had the hashtags \#devam or \#tamam in their profile description as supporting or opposing Erdo\u{g}an. Lastly, users who had the hashtag \#RTE (Recep Tayyip Erdo\u{g}an) in their profile description were labeled as pro-Erdo\u{g}an. While providing a political party name as a part of twitter user profile is a strong indication of supporting the respective party, we manually checked all extracted names to ensure the correctness of labels. For instance, we found that some users expressed that they are against a particular party in their user name instead of supporting it. Therefore, whenever we suspected that keywords we used for labeling were not indicative of their political view, we manually investigated the accounts and removed their labels if their political views were unclear. The total number of manually labeled users were as follows:  
\begin{center}
\begin{tabular}{|l|r|}  
\hline
\textbf{Supporters} &  \textbf{Labeled Users}  \\  \hline
pro-Erdo\u{g}an & 1,777 \\ \hline
anti-Erdo\u{g}an w/out party affiliation & 2,134 \\ \hline
pro-CHP & 833 \\ \hline
pro-IYI & 900 \\ \hline
pro-HDP & 357 \\ \hline
\textbf{Total} & 3,866 \\
\hline
		\end{tabular}
\end{center}
\textbf{(2) Label Propagation:} Label propagation automatically labels users based on the tweets that they retweet \cite{darwish2017predicting,darwish2018scotus}. The intuition behind this method is that users that retweet the same tweets most likely share the same stances on the topics of the tweets. Given that many of the tweets in our collection were actually retweets or duplicates of other tweets, we labeled users who retweeted 10 or more tweets that were authored or retweeted by the pro- or anti- groups and no retweets from the other side as pro- or anti- respectively. We iteratively performed such label propagation 11 times, which is when label propagation stopped labeling new accounts. By the last iteration, we had labeled 652,729 users of which 279,181 were pro-Erdo\u{g}an and posted 28,050,613 tweets, and 373,548 were anti-Erdo\u{g}an and posted 31,762,639 tweets. To ensure labeling accuracy, we manually and independently labeled 100 users from each of both the pro and anti groups, and found that label propagation matched manual labeling for 191 users, 1 user label was clearly wrong, and we could not decide on the stance of 8 remaining users due to lack of political tweets. 
\paragraph{\textbf{Timeline Data Crawling:}}\label{sec_timeline_crawling} 
On Dec. 28, 2018, we started crawling the timelines of 86,116 and 81,963 pro and anti users respectively using Tweepy API\footnote{http://www.tweepy.org/}. Twitter typically allows the crawling of the latest 3,200 tweets for a user. Depending on how active each user is, 3,200 tweets can cover days, months, or years. We also excluded all non-Turkish tweets. In all, we collected 98,700,529 and 115,047,039 tweets from pro and anti groups respectively, with some of the tweets dating back to 2013.


\section{Data Preparation}
Due to the informal nature of Twitter, tweets commonly have grammatical and spelling errors.  Further, 
Twitter users frequently use emojis, emoticons, hashtags, media links, and other non-alphabetic characters. 
Thus, we performed the following pre-processing steps for the tweets:\\ 
-- Case folding, where we lower-cased letters.\\
-- Removal of all links, hashtags, and user mentions.\\
-- Removal of all non-letter characters and punctuations. \\ 
-- Replacement of all numbers to the word ``number''. \\

Since Turkish is an agglutinative language, approximately 60 surface forms can be generated from a single root without changing the POS tag of the word~\cite{oflazer2018turkish}.  For example, the word ``okullar{\i}m{\i}zdan'' (``from our schools'') consists of the morphemes okul+lar+{\i}m{\i}z+dan (``school+plural-marker+our+from) and is derived from the root ``okul'' (``school''). Having suffixes not only can hurt the performance but also can cause the word embedding model to generate many different word vectors for one stem. To avoid both of these, we stemmed all words using the Snowball Turkish Stemmer \cite{ccilden2006stemming}. Lastly, we converted Turkish specific letters to English letters that are closest in shape (e.g.,. \u{g} $\rightarrow$ g). 
Given the cleaned and stemmed tweets, we trained 4 separate word embeddings, namely ED-pro, ED-anti, TD-pro, and TD-anti for the pro or anti tweets of both the ED and TD datasets.


\section{Our Approach}

Our approach extends the work of Garg et al.~\cite{garg2018word}, which showed that word embeddings (WEs) are able to capture gender and racial stereotypes by comparing embeddings that are built on different texts 
to understand how a given term is defined semantically by different groups of people. 
We initially train WEs using the different datasets separately. 
We use fastText~\cite{joulin2016bag} to train the embeddings.  FastText is advantageous, because it represents words as a ``bag of character n-grams'' allowing to effectively overcome the agglutinative nature of Turkish, where suffixes indicate morphological and syntactic functions \cite{bojanowski2017enriching} and spelling mistakes are very common in tweets. We train a skip-gram model with default parameters (learning rate = 0.05; vector size = 100; epochs  = 5). 
For a given term $t$, we then find the nearest terms in each embedding space. 
Then we inspect the nearest terms at each ranking and compare their semantics and sentiment. For instance, we expect that words with positive sentiment are closer to the word ``Erdo\u{g}an'' in WE spaces built on tweets of pro-Erdo\u{g}an group, and expect the opposite for tweets of anti-Erdo\u{g}an group. We do not compare the actual distances of words since we use different WE spaces; instead, we compare the ranks of words. We compiled a list of 280 positive and negative sentiment adjectives. Next, given an entity of interest, we use fastText to obtain the top 2,000 Nearest Neighbors (NNs).
We singled out two types of NNs. First, we identified ones that appear in our list of adjectives. Due to common spelling errors, we allowed matches that were within 1 Levenshtein edit distance away. Second, we captured NNs that subsume entity names. This is important, because users often use names of entities with identifiers (e.g., ``liderim Erdo\u{g}an'' (my leader Erdo\u{g}an); and though we removed hashtags, many concatenated words were prevalent (e.g. ``oyumErdo\u{g}ana'' (my vote is for Erdo\u{g}an). 
Finally, we manually inspected the filtered NNs to ascertain their sentiment or semantics. 

\section{Qualitative Analysis}
We compared the sentiment of different groups towards three types of categories, namely: 1) names of politicians; 2) political issues; and 3) lifestyle matters. 

\paragraph{\textbf{Sentiment Towards Politicians}}\label{sec_politicians}
We compared the sentiment towards 5 prominent politicians across our four WEs (both pro and anti for ED and TD datasets).  The politicians were: Recep Tayyip Erdo\u{g}an (AkParti Leader), Devlet Bah\c{c}eli (MHP leader), Kemal K{\i}l{\i}\c{c}daro\u{g}lu (CHP leader), Meral Ak\c{s}ener (IYI Party leader), and Selahattin Demirta\c{s} (HDP leader). We used fastText to obtain 2,000 Nearest Neighbors (NN) to every politician's name in all four WEs.  \textbf{Tables~\ref{tab1e_politician_timeline}, \ref{tab1e_politician_election}} list NNs matching our sentiment adjectives or subsuming politicians' names with their rank for the ED and TD datasets, respectively. Though some words do not have universally negative sentiment, we considered their sentiment in the context of Turkish politics. For instance, ``Americanist'' usually refers to a politician who puts America's benefit first -- as opposed to Turkey's interest. Insults without a direct English translation are just translated as ``insult''. 
As expected for Erdo\u{g}an, words with positive sentiments ranked higher in TD-pro, whereas negative-sentiment words ranked higher in TD-anti. For example, phrases with possessive suffices such as ``my/our president'' and ``our leader'' appeared more than 40 times in TD-pro in high ranks, while ``my president'' appears once (at rank 1066) in TD-anti.  This suggests that pro-Erdo\u{g}an users consider him as a great leader, so they tend to use phrases like ``our leader'', ``world leader'', ``leader of the century'', and ``custodian of the republic''. In contrast, anti-Erdo\u{g}an users described him mostly as a ``dictator'' and used many insulting phrases. Bah\c{c}eli, who supported Erdo\u{g}an in the election, received similar treatment.
Our approach captures the subtle nuance of alliances in polarized political atmospheres. Ak\c{s}ener and Bah\c{c}eli are both nationalists. However, prominent TD-pro terms branded Ak\c{s}ener as the ``enemy of the nationalists'' and ``inconsistent'', while branding Bah\c{c}eli as a ``leader'' (``leader of Turks'', ``the only leader'', etc.). This may indicate that positions towards them are shaped more by political alliances than ideology. It is noteworthy that Ak\c{s}ener defected from MHP to establish IYI Party.  Though sentiments towards Ak\c{s}ener and K{\i}l{\i}\c{c}daro\u{g}lu were consistently negative in TD-pro, sentiments were mixed in TD-anti, with more negative-sentiment words.
Top ranking TD-pro terms identified Demirta\c{s} 
as a criminal (e.g., ``terrorist'', ``PKK supporter'', ``murderer'', ``baby killer''), while Anti-Erdo\u{g}an users viewed him as a victim. 
NNs from ED-pro and ED-anti WEs (Table~\ref{tab1e_politician_election}) were consistent in direction with TD WEs for Erdo\u{g}an, Bah\c{c}eli, and K{\i}l{\i}\c{c}daro\u{g}lu. Certain terms, however, rising in rank.  For example, for Erdo\u{g}an the rank of ``dictator'' increased from 490 to 143, and the rank of "evil born" rose from 239 to 81.  Similarly for Bah\c{c}eli, the rank of ``deep state man'' climbed from 372 to 153.  We observed notable changes for Ak\c{s}ener and Demirta\c{s}.  For Ak\c{s}ener, ED-anti sentiment was mostly positive -- compared to mixed sentiment from TD-anti. Conversely, ED-anti sentiment was mixed for Demirta\c{s} -- compared to mostly positive from TD-anti.  
This suggests that users became more polarized in the lead up to the election, and they showed less sympathy towards opposing politicians. 
Also, it seems that Anti-Erdo\u{g}an users were mostly agreeing on being against Erdo\u{g}an, but were not necessarily supporting each other.
\begin{table*}[ht]
\caption{NNs for given politician names in Timeline datasets. \textit{Positive} sentiment words are \textit{italicized}, and \textbf{negative} sentiment words are \textbf{bolded}. Each term is followed by its rank (in parenthesis). If a term appeared more than once, we list its highest rank. }\label{tab1e_politician_timeline}

\begin{tabular}{| p{1.5cm} | l |  p{14.2 cm} |}  \hline
		\textbf{Politician} & \textbf{Dataset} & \textbf{closest identifiers} \\ \hline
		\multirow{2}{*}{\bf  {Erdo\u{g}an}} & \textbf{TD-pro} &
    		\emph{my/our president/prime minister} (20, 37x); \emph{our Erdo\u{g}an} (308, 833);  leader  (97,7x);  \emph{independent president}  (117)   \emph{leader (w/ positive sentiment)} (96, 3x);  \emph{(my) military commander in chief}  (312, 5x)  \emph{president of the people} (409) \textbf{dictator} (704, 6x) \textbf{insult} (706,3x)  \emph{the man of the people} (739, 2x)  \textbf{pkk supporter} (758) \emph{our party  leader} (802)  \emph{our leader} (867,3x) \textbf{actor} (900) \textbf{murderer} (953) \emph{revolutionist} (1015) \emph{world leader} (1286,2x) \emph{leader of the century} (1407, 3x)  \emph{custodian of the republic} (1410,2x) \\ \cline{2-3}
    		 & \textbf{TD-anti} &   \emph{independent president} (239); \textbf{evil born} (268); akp supporter(454);  \textbf{dictator} (490,6x); \textbf{insult} (333, 6x); leader (506,6x);  opponent  (604); \textbf{neonazi fuhrer} (630);  \textbf{sultan} (827); \textbf{murderer}  (953,2x); \emph{my president} (1066);  \textbf{cruel} (1104); \emph{person who makes Israel kneel down} (1196);  \textbf{lyer} (1463); \textbf{making unbearable} (1596);       \textbf{traitor} (1751);   \textbf{terrorist} (1832);         Muslim(1886)     \\ \hline  \hline
		 \multirow{2}{1.75cm}{\bf {K{\i}l{\i}\c{c}daro\u{g}lu}} & \textbf{TD-pro} &  \textbf{slanderous} (120)                 \textbf{dictator} (170,5x)            \textbf{insult} (181,4x)        \textbf{political trash} (279)      \emph{hope of the people} (289)        \textbf{political pervert} (303)     \textbf{inconsistent} (353,3x)         \textbf{immoral} (385)         \textbf{enemy of the women} (471)     \textbf{absurd} (572)        \emph{leader of the Muslims} (692) \textbf{shameless} (941,1748); \textbf{lyer} (1190,3x) \textbf{barefaced} (1366)
		    \\ \cline{2-3}
		 & \textbf{TD-anti} &   \textbf{immoral} (152);       \emph{hope of the people} (185);           leader(233,1735); \emph{our prime minister} (287);          \emph{the fear of immoral people} (387);       \emph{inconsistent} (866)        \\  \hline  \hline
		 \multirow{2}{1.75cm}{\bf   {Ak\c{s}ener}} & \textbf{TD-pro} &   leader(62)     \textbf{weird} (83)       \textbf{shameless} (119);                \emph{our hope} (195);     \textbf{FET\"{O} supporter} (224, 2x)              \textbf{malicious} (319);      \textbf{enemy of the nationalists} (339);         \textbf{insult} (869);        \textbf{the voice of the people(1328)}        \textbf{inconsistent} (1813,2x)  \emph{the candidate of the people} (1921)  \\ \cline{2-3}
		 & \textbf{TD-anti} &  \textbf{FET\"{O} supporter} (83);           leader(165,3x); \emph{the candidate of the people} (267);      \emph{the voice of the people} (277);         \emph{our hope} (282)        \emph{antidote} (315);         \emph{the sun of the goodness} (388);        \emph{our leader} (427)        \emph{fearless} (646);     \textbf{shameless} (1133);        \emph{the last hope} (1317)    \\ \hline  \hline
		 \multirow{2}{1.75cm}{\bf   {Bah\c{c}eli}} & \textbf{TD-pro} & \emph{leader (w/ positive sentiment)} (84,3x)     leader(89,12x); \emph{our leader} (104);         \emph{wise} (181);          \emph{statesman} (201), \emph{the leader of Turks} (294);            the leader of nationalist movement(557);     \emph{the voice of personnel of supportive service} (791);     \emph{the only leader} (797);             \emph{the hope of the imprisoned} (882);       \textbf{loser} (1146,1464)            \\ \cline{2-3}
		 & \textbf{TD-anti} &  \textbf{deep state's man} (372); leader(1323,2x)         \\ \hline  \hline
		 \multirow{2}{1.75cm}{\bf   {Demirta\c{s}}} & \textbf{TD-pro} &  \textbf{PKK supporter} (22)            father(59)     \textbf{murderer} (75,2x)       \textbf{terrorist} (81,24x)       fascist(137)     dishonor(219)   baby killer(225,2x)     saz player (349)     \\ \cline{2-3}
		 & \textbf{TD-anti} &  leader(31,4x)        \emph{oppressed(74)}          father(108)          \emph{ our candidate(114)};            \emph{victim(214)}           \emph{friend(266,4x)};       \emph{democrat} (783);      \textbf{insult} (1882)      \\ \hline 
	\end{tabular}
\end{table*}

\begin{table*}[ht]
\caption{NNs for given politician names in the Election datasets. \textit{Positive} sentiment words are \textit{italicized}, and \textbf{negative} words are \textbf{bolded}. Each term is followed by its rank (in parenthesis). If a term appeared more than once, we list its highest rank.}\label{tab1e_politician_election}
\begin{tabular}{| p{1.7cm} | l |  p{14.2 cm} |}  \hline
		 \textbf{Politician} & \textbf{Dataset} & \textbf{closest identifiers} \\ \hline
		\multirow{2}{*}{\bf  Erdo\u{g}an} & \textbf{ED-pro} &  \emph{my/our president (44, 30x)} leader (74,5x); \textbf{communist} (252);  \emph{my/our leader} (279, 2x); \textbf{insult} (317); \textbf{dictator} (322, 2x); \emph{my Erdo\u{g}an(443, 2x)} reis (554);	\emph{leader of the people (604)}  \emph{commander-in-chief} (620, 2x); \emph{world leader} (627); \textbf{americanist} (674);  opponent (901); \emph{anti-Zionist} (993)  \\ \cline{2-3}
		 & \textbf{ED-anti} &  \textbf{evil born} (81); \textbf{insult} (85, 4x); \textbf{dictator} (143,9x); \textbf{dog} (181);  \textbf{little leader} (226); \textbf{bullyboy} (268); opponent (313, 3x); \textbf{thief} (408,2x); \textbf{propagandist} (418); \textbf{idiot} (420, 4x);  \textbf{muppet} (442, 736); \emph{our president} (551);   leader (612), \textbf{the head of FET\"{O}} (645); \emph{independent president} (697); \textbf{American} (702), \textbf{traitor} (737), \textbf{lyer} (759, 2x); \textbf{fake} (784); \textbf{despot} (795);  \emph{leader w/ positive sentiment} (860); conservative (877); Ecevit supporter (928);  \\ \hline  \hline
		 \multirow{2}{2.5cm}{\bf   K{\i}l{\i}\c{c}daro\u{g}lu} & \textbf{ED-pro} &   \textbf{inconsistent} (363,5x);  \textbf{empty} (439)   \textbf{idiot} (479,2x);       \textbf{lyer} (526,3x);            \textbf{dog} (1466)        chief (1593)   \\ \cline{2-3}
		 & \textbf{ED-anti} &  \textbf{ inconsistent} (248,2x)     leader (273, 5x)         \emph{leader with positive sentiment (1092)};        \textbf{dumb} (1233) \\ \hline  \hline
		 \multirow{2}{1.75cm}{\bf  Ak\c{s}ener}  & \textbf{ED-pro} &   \textbf{insult}  (33); \textbf{candidate of FET\"{O}} (75);     \textbf{unreliable} (492)     \textbf{inconsistent} (1061);       \textbf{traitor} (1066);         \emph{great} (1280)      \\ \cline{2-3}
		 & \textbf{ED-anti} &  \emph{the candidate of the people} (254, 2x), \textbf{insult} (462);\emph{ brave} (485, 3x)             \emph{precious} (966), leader (1329)   \emph{great} (1819)     \\ \hline  \hline
		 \multirow{2}{1.75cm}{\bf   Bah\c{c}eli} 
		 & \textbf{ED-pro} &  \emph{the hope of imprisoned} (33);       leader (42,4x); sensitive (150,3x)      \emph{precious} (390,2x); \emph{great}    (533,2x)     \emph{wise} (673)           \textbf{confused} (1027)        \emph{problem solver} (1107)    bald (1299); \emph{respected} (1715)  
   \\ \cline{2-3}
		 & \textbf{ED-anti} &  \textbf{deep state's man} (153); \textbf{FET\"{O} supporter} (836)      \textbf{insult}  (1025); leader (1173,1992)     \emph{successful} (1507)   \\ \hline  \hline
		 \multirow{4}{1.75cm}{\bf   Demirta\c{s}} &  \textbf{ED-pro} & \textbf{terrorist} (45, 32x); citizen (111); \emph{friend} (111); \textbf{murderer} (411, 4x); \textbf{insult}  (457)          \textbf{shameless} (565, 3x);  \textbf{barefaced} (1223)    \textbf{idiot} (1597);      \textbf{dictator} (1939)        \textbf{dishonored} (1946)        \\ \cline{2-3}
		 & \textbf{ED-anti} &   \emph{our candidate} (117,2x);         \textbf{murderer} (127,2x)       \emph{my president} (157); citizen (199,3x); \emph{comrade} (199); \emph{friend} (199,4x); \textbf{insult}  (276); \textbf{terrorist} (419,10x);        \textbf{confused} (445)             liberal (888)  \textbf{mean} (997); \emph{democrat}    (1006); \textbf{idiot} (1071); \emph{fare} (1162)     \textbf{dictator} (1250); free(1589)      \\ \hline  
		\end{tabular}
\end{table*}

\paragraph{\textbf{Stance Towards Political Issues}}\label{sec_political_issues} 
We focused on three political issues that were frequently discussed on Turkish media, namely: 1) Syrian refugees in Turkey; 2) the Kurdish YPG armed-group in Syria, which is supported by the USA and declared as a terrorist organization by Turkey; 
and 3) FET\"{O} which is a Turkish organization, led by Fethullah G\"{u}len, who is living in Pennsylvania, and which Turkey has labeled a terrorist organization for its alleged role in 2016 attempted coup. The two latter topics are issues of contention between the US and Turkey. 
We polled all WEs models with the terms ``Suriyeli'' (Syrian), ``YPGli'' (YPG supporter) and ``FET\"{O}''. 
\textbf{Table~\ref{tab1e_issues}} lists NNs expressing sentiment on all topics from all four WEs.   Regarding Syrians, we can see that many negative words (e.g., ``murderer'', ``terrorist'', ``parasite'') are remarkably present in TD-anti and ED-anti, while being noticeably scarce in TD-pro and ED-pro.  
This is consistent with anti-refugee stances of the anti-Erdo\u{g}an camp, as verbalized by CHP and IYI party leaders\footnote{e.g., a tweet of Meral Ak\c{s}ener \url{twitter.com/meral\_Ak\c{s}ener/status/1062339295871152129}}. Positive sentiment from the pro-Erdo\u{g}an camp is interesting given the MHP's traditional anti-refugee stance.  It could be the result of: AKParti supporters outnumbering MHP supporters; MHP assuming positions closer to AKParti due to their alliance; or MHP's base being less different on the issue compared to the party's leadership.  This requires more investigation. Regarding YPG and FET\"{O}, we observe that both groups share similar sentiment regardless of the elections, with negative sentiment NNs ranking higher in pro-Erdo\u{g}an camp than anti-Erdo\u{g}an camp. 

\begin{table*}[ht]
\caption{NNs for the given political issues in Timeline and Election datasets.  \textit{Positive} terms are \textit{italicized} and \textbf{negative} terms are \textbf{bolded}.  Terms are followed by rank and frequency. If a term appeared more than once, we list its highest rank. }\label{tab1e_issues}
\begin{tabular}{| p{1.5cm} | l |  p{14.2 cm} |}  \hline
	    \textbf{Politician}	  & \textbf{Dataset} & \textbf{closest identifiers} \\ \hline
		\multirow{4}{*}{\bf  {Syrian}}  & \textbf{TD-pro} & Muslim (1784) 	\\ \cline{2-3}
    		 & \textbf{TD-anti} &  Arab (33), foreigner(44), \textbf{murderer} (64), Muslim (407,1359), \textbf{terrorist} (1274)   	\\ \cline{2-3}
    		 & \textbf{ED-pro} &  sharing the same religion (534), modern (1404), \textbf{lousy} (1688) 	\\ \cline{2-3}
    		  & \textbf{ED-anti} &  \textbf{swindler} (358), \textbf{dishonorable} (1413), \textbf{parasite} (1541), democrat (1986)
    		 \\ \hline  \hline
        	\multirow{4}{*}{\bf  {YPG supporter}}  & \textbf{TD-pro} &  \textbf{PKK supporter} (14) \textbf{terrorist} (40, 39x)
		\\ \cline{2-3}
    		 & \textbf{TD-anti} &   \textbf{terrorist} (615, 13x), \textbf{traitor} (1326)    	\\ \cline{2-3}
    		 & \textbf{ED-pro} &  \textbf{terrorist} (7, 30x),   \textbf{dishonorable} (85,17x), \textbf{insult} (156,1590), \textbf{traitor} (301, 5x), \textbf{swindler} (314), kemalist (460, 2x), communist (841), right leaning (1539) jealous (1562), \textbf{pyschopath} (1668) \textbf{mean} (1861) socialist (1873) \textbf{untrustworthy} (1896) 	\\ \cline{2-3}
    		  & \textbf{ED-anti} &   \textbf{terrorist} (69, 36x), \textbf{traitor} (335, 1,553), \textbf{dishonorable} (495,15x), \textbf{insult} (1,014, 1,521) leftist (1,392), \textbf{murderer} (1,886), \textbf{bigot} (1,989)
    		 \\ \hline  \hline
		 \multirow{4}{*}{\bf  {FET\"{O}}}  & \textbf{TD-pro} & 
	    \textbf{coup organizer} (26,7x) mason (81), nationalist (82), \textbf{traitor} (451, 2x), \textbf{americanist} (517), good (536), \textbf{puppet} (740, 2x) idiot (999) Gezi protester (1001)
		\\ \cline{2-3}
    		 & \textbf{TD-anti} & \textbf{traitor} (162, 4x), \textbf{coup organizer} (253, 2x), \textbf{americanist} (566), \textbf{pkk supporter} (889), \textbf{terrorist} (1008, 8x)  kemalist (1104) \textbf{criminal} (1261)   	\\ \cline{2-3}
    		 & \textbf{ED-pro} &  \textbf{insult} (67,219), \textbf{puppet} (82), Nurist\footnote{a religious group in Turkey}, \textbf{evil} (161) \textbf{dishonorable} (192), \textbf{coup organizer} (193) \textbf{traitor} (452, 3x) \textbf{terrorist} (624, 9x), PKK supporter (736, 2x)	\\ \cline{2-3}
    		  & \textbf{ED-anti} &  \textbf{terrorist} (376, 19x), \textbf{dishonorable} (1072), \textbf{shameless} (1384) \textbf{traitor} (1418), pkk supporter (1418) \textbf{not virtuous} (1697)
    		 \\ \hline
	\end{tabular}
\end{table*}
\begin{table*}[ht!]
\caption{NNs for lifestyle words for pro- and anti-Erdo\u{g}an groups. Terms are followed by rank andfrequency. If a term appeared more than once, we list its highest rank.}\label{tab1_non_political}
\begin{tabular}{| p{1.5cm} | l|  p{14.2 cm} |}  \hline
        \textbf{Term} & \textbf{Dataset} & \textbf{Closest examples} \\ \hline
		\multirow{2}{1.75cm}{\bf icecek (drink)} & \textbf{TD-pro} & soft drink (3, 8x), lemonade (22), buttermilk (24), juice (25), coffee (26), cigarette (28), milkshake (31), non-alcoholic (47), macchiato (55), bleach (56, 87), wine (59), kefir (60), alcohol (67), carbonated (74,2x), soda (82,95)    \\ \cline{2-3}
		 & \textbf{TD-anti} &  buttermilk (6, 2x), cigarette (12, 3x), soft drink (16, 3x), different alcoholic beverages (19, 10x), kefir (23), redbull (26,2x), sahlep (43), coffee (60, 2x), milkshake (65), antacid (68), medication (74), iced tea (85), non-alcoholic (91), lemonade(95, 2x), americano (98)  \\ \hline \hline
		 	\multirow{2}{1.75cm}{\bf Yemek (meal/to eat)} & \textbf{TD-pro} & breakfast(2, 20x), cake (3), iftar (8, 4x), soup (19, 3x), sandwich (23, 2x), cig kofte (46), tirit (52, 2x), paste (57), meatballs \& bread (60), roasted meat (67), sahoor (70), pastry (74, 2x), simit\&tea\&cheese (89),  sujuk (91), meatballs (92), rice(93),      \\ \cline{2-3}
		 & \textbf{TD-anti} &  breakfast(5, 3x), meal \& desert(9), meal \& tea(12), sandwich (16, 2x), meal \& coffee (21), bean \& rice \& pickles (23), meal \& wine (25), without meat (34), beans \& rice (35, 3x), hamburger (37), soup-meal-desert (48), sujuk (51), pita-pizza-soup (52), chips (56), spinach (59), fillet steak \& rice (61), youghurt (63), fat (67), pastry (70), celery (74), meatballs (78, 2x), pasta \& potato (79), pudding (84), chicken (85), roasted (86), baklava-borek (89), bechamel sauce (96), egg (98)     \\ \hline \hline
		  	\multirow{2}{1.75cm}{\bf tatil (vacation)} & \textbf{TD-pro} & vacation village (1, 2x), fun (36, 5x), sea \& hotel (48), snowman (75), Marmaris (82)     \\ \cline{2-3}
		 & \textbf{TD-anti} &   sunbathing (65, 6x), Rhodes island (75), Maldives (87)  \\ \hline
		\end{tabular}
\end{table*}
\paragraph{\textbf{Lifestyle Issues}}\label{sec_nonpolitical_issues} 
Lastly, we examined lifestyle related issues in the pro- and anti-Erdo\u{g}an camps for the Timeline Election dataset. We looked at 3 lifesyle issues, namely: drinks (icecek), food (yemek), and vacations (tatil).  Unlike political leaders and political issues, we did not expect to see positive or negative sentiment words in the top NNs.  Thus, we manually examined the top 100 NNs to identify terms that may indicate different lifestyle choices.  Since elections typically have little bearing on what people eat or drink, we only queried the TD-pro and TD-anti WE models.  
\textbf{Table~\ref{tab1_non_political}} shows the rank of such terms in the TD-pro and TD-anti an groups. We grouped similar items to improve readability. For instance, "Pepsi" and "Cola" are both put under ``soft drinks''.  Our WE models captured the disparity in life style of the two groups. Upon inspecting NNs of the word ``drink'',  alcoholic beverages ranked 59 and 67 in TD-pro, while they appeared 10 times in TD-anti with higher ranks. Similarly for food, the words ``iftar'' and ``sahoor'', which are closely associated with Muslim fasting, appeared in TD-pro but not in TD-anti.  Additionally, we also noticed many traditional dishes appearing in TD-pro such as \c{c}i\u{g} kofte (raw meatball), tirit (meat and organ dish), simit (baked rolls), and sujuk (Turkish sausages). Conversely, in TD-anti, multiple fast-food items such as burgers, pizza, and chips featured prominently along with other non-Turkish cuisines such as bechamel sauce and pudding. Such are consistent with election results. Tirit is popular in cities such as Konya\footnote{http://konya.com.tr/en/portfolio-item/tirit-kebabi/}, Samsun\footnote{https://www.samsunkavak.bel.tr/Sayfa/kaz-tirit-yemegi}, and Sanliurfa\footnote{https://www.gundemsanliurfa.com/sanliurfa-yemekleri/urfa-tirit-yemegi-h10.html}, and cig kofte (raw meatball) is popular in SanliUrfa\footnote{https://en.wikipedia.org/wiki/\%C3\%87i\%C4\%9F\_k\%C3\%B6fte}. Voters in Konya, Samsun and Sanliurfa, gave Erdo\u{g}an 74\%, 61\%, and 65\% of the votes respectively\footnote{https://secim.haberler.com/2018/}. Fast-food and foreign cuisine are more popular in large cities. Erdo\u{g}an and his party seem to fair much better in elections in rural areas compared to cities. 
For ``vacation'', we see that ``sun-bathing'' appears multiple times along with foreign destinations in TD-anti, yet completely absent for TD-pro. This arguably indicates divergent levels of religiosity between TD-pro and TD-anti groups.  

\section{Conclusion}

In this work, we investigated polarization concerning the 2018 Turkish elections through tweets. We collected 108M tweets posted during the period leading up to the election and we used a semi-supervised method to label 652.7K users as pro- or anti-Erdo\u{g}an, posting 59.8M tweets in total. Subsequently, we randomly selected 168K labeled users and crawled their timelines, collecting 213.7M tweets in total, covering the period before and after the election. 
We built 4 different word embedding models using tweets of pro- and anti-Erdo\u{g}an groups separately from both datasets. Using word embeddings, we explored polarization between the two groups on different topics during different time intervals. By looking at the neighboring terms to entities of interest in an embedding space, we were able to capture the sentiment of users towards an entity. We were also able to capture political attitudes and lifestyle choices of users. Our analysis shows that: (1) there was strong polarization towards political leaders and it further increased during the election; (2) users were clearly polarized on certain political issues, such Syrian refugees; (3) polarization was correlated with specific lifestyle issues such as what people eat; and (4) despite the polarization, both anti- and pro-Erdo\u{g}an groups agreed on topics pertaining to security and democracy.



\bibliographystyle{splncs04} 
\bibliography{refs}

\begin{thebibliography}{10}
\providecommand{\url}[1]{\texttt{#1}}
\providecommand{\urlprefix}{URL }
\providecommand{\doi}[1]{https://doi.org/#1}

\bibitem{bojanowski2017enriching}
Bojanowski, P., Grave, E., Joulin, A., Mikolov, T.: Enriching word vectors with
  subword information. Transactions of the Association for Computational
  Linguistics  \textbf{5},  135--146 (2017)

\bibitem{borge2015content}
Borge-Holthoefer, J., Magdy, W., Darwish, K., Weber, I.: Content and network
  dynamics behind egyptian political polarization on twitter. In: Proceedings
  of the 18th ACM Conference on Computer Supported Cooperative Work \& Social
  Computing. pp. 700--711. ACM (2015)

\bibitem{caliskan2017semantics}
Caliskan, A., Bryson, J.J., Narayanan, A.: Semantics derived automatically from
  language corpora contain human-like biases. Science  \textbf{356}(6334),
  183--186 (2017)

\bibitem{ccilden2006stemming}
{\c{C}}ilden, E.K.: Stemming turkish words using snowball (2006)

\bibitem{conover2011political}
Conover, M., Ratkiewicz, J., Francisco, M.R., Gon{\c{c}}alves, B., Menczer, F.,
  Flammini, A.: Political polarization on twitter. Icwsm  \textbf{133},  89--96
  (2011)

\bibitem{bias_assim}
Dandekar, P., Goel, A., Lee, D.T.: Biased assimilation, homophily, and the
  dynamics of polarization. Proceedings of the National Academy of Sciences
  \textbf{110}(15),  5791--5796 (2013). \doi{10.1073/pnas.1217220110},
  \url{https://www.pnas.org/content/110/15/5791}

\bibitem{darwish2018scotus}
Darwish, K.: To kavanaugh or not to kavanaugh: That is the polarizing question.
  arXiv preprint arXiv:1810.06687  (2018)

\bibitem{darwish2017predicting}
Darwish, K., Magdy, W., Rahimi, A., Baldwin, T., Abokhodair, N.: Predicting
  online islamophopic behavior after\# parisattacks. The Journal of Web Science
   \textbf{3}(1) (2017)

\bibitem{darwish2017improved}
Darwish, K., Magdy, W., Zanouda, T.: Improved stance prediction in a user
  similarity feature space. In: Proceedings of the 2017 IEEE/ACM International
  Conference on Advances in Social Networks Analysis and Mining 2017. pp.
  145--148. ACM (2017)

\bibitem{darwish2019unsupervisedStance}
Darwish, K., Stefanov, P., Aupetit, M.J., Nakov, P.: Unsupervised user stance
  detection on twitter. arXiv preprint arXiv:1904.02000  (2019)

\bibitem{duan2012graph}
Duan, Y., Wei, F., Zhou, M., Shum, H.Y.: Graph-based collective classification
  for tweets. In: Proceedings of the 21st ACM international conference on
  Information and knowledge management. pp. 2323--2326. ACM (2012)

\bibitem{garg2018word}
Garg, N., Schiebinger, L., Jurafsky, D., Zou, J.: Word embeddings quantify 100
  years of gender and ethnic stereotypes. Proceedings of the National Academy
  of Sciences  \textbf{115}(16),  E3635--E3644 (2018)

\bibitem{garimella2018quantifying}
Garimella, K., Morales, G.D.F., Gionis, A., Mathioudakis, M.: Quantifying
  controversy on social media. ACM Transactions on Social Computing
  \textbf{1}(1), ~3 (2018)

\bibitem{giatsoglou2017sentiment}
Giatsoglou, M., Vozalis, M.G., Diamantaras, K., Vakali, A., Sarigiannidis, G.,
  Chatzisavvas, K.C.: Sentiment analysis leveraging emotions and word
  embeddings. Expert Systems with Applications  \textbf{69},  214--224 (2017)

\bibitem{guerra2013measure}
Guerra, P.C., Meira~Jr, W., Cardie, C., Kleinberg, R.: A measure of
  polarization on social media networks based on community boundaries. In:
  Seventh International AAAI Conference on Weblogs and Social Media (2013)

\bibitem{joulin2016bag}
Joulin, A., Grave, E., Bojanowski, P., Mikolov, T.: Bag of tricks for efficient
  text classification. arXiv preprint arXiv:1607.01759  (2016)

\bibitem{kumar2018community}
Kumar, S., Hamilton, W.L., Leskovec, J., Jurafsky, D.: Community interaction
  and conflict on the web. In: Proceedings of the 2018 World Wide Web
  Conference on World Wide Web. pp. 933--943. International World Wide Web
  Conferences Steering Committee (2018)

\bibitem{kutlu2018devam}
Kutlu, M., Darwish, K., Elsayed, T.: Devam vs. tamam: 2018 turkish elections.
  arXiv preprint arXiv:1807.06655  (2018)

\bibitem{magdy2016isisisnotislam}
Magdy, W., Darwish, K., Abokhodair, N., Rahimi, A., Baldwin, T.: \#
  isisisnotislam or\# deportallmuslims?: Predicting unspoken views. In:
  Proceedings of the 8th ACM Conference on Web Science. pp. 95--106. ACM (2016)

\bibitem{magdy2016failedrevolutions}
Magdy, W., Darwish, K., Weber, I.: \# failedrevolutions: Using twitter to study
  the antecedents of isis support. First Monday  \textbf{21}(2) (2016)

\bibitem{homophily}
McPherson, M., Smith-Lovin, L., Cook, J.M.: Birds of a feather: Homophily in
  social networks. Annual Review of Sociology  \textbf{27}(1),  415--444
  (2001). \doi{10.1146/annurev.soc.27.1.415},
  \url{https://doi.org/10.1146/annurev.soc.27.1.415}

\bibitem{morales2015measuring}
Morales, A., Borondo, J., Losada, J.C., Benito, R.M.: Measuring political
  polarization: Twitter shows the two sides of venezuela. Chaos: An
  Interdisciplinary Journal of Nonlinear Science  \textbf{25}(3),  033114
  (2015)

\bibitem{homophily_twitter}
Mousavi, R., Gu, B.: The effects of homophily in twitter communication network
  of u.s. house representatives: A dynamic network study. SSRN Electronic
  Journal  (01 2015). \doi{10.2139/ssrn.2666052}

\bibitem{oflazer2018turkish}
Oflazer, K., Sara{\c{c}}lar, M.: Turkish Natural Language Processing. Springer
  (2018)

\bibitem{weber2013secular}
Weber, I., Garimella, V.R.K., Batayneh, A.: Secular vs. islamist polarization
  in egypt on twitter. In: Proceedings of the 2013 IEEE/ACM International
  Conference on Advances in Social Networks Analysis and Mining. pp. 290--297.
  ACM (2013)

\bibitem{wong2013quantifying}
Wong, F.M.F., Tan, C.W., Sen, S., Chiang, M.: Quantifying political leaning
  from tweets and retweets. In: Seventh International AAAI Conference on
  Weblogs and Social Media (2013)

\bibitem{wong2016quantifying}
Wong, F.M.F., Tan, C.W., Sen, S., Chiang, M.: Quantifying political leaning
  from tweets, retweets, and retweeters. IEEE transactions on knowledge and
  data engineering  \textbf{28}(8),  2158--2172 (2016)

\end{thebibliography}
\end{document}